\documentclass[lettersize,journal]{IEEEtran}

\PassOptionsToPackage{table}{xcolor}
\usepackage[nocompress]{cite}
\usepackage{amsmath,amssymb}
\usepackage{booktabs}
\usepackage{graphicx}
\usepackage{xcolor}
\usepackage{listings}
\usepackage{algorithm}
\usepackage{algpseudocode}
\usepackage{balance}
\usepackage{url}
\usepackage{enumitem}
\usepackage{array}
\usepackage{tabularx}
\usepackage{pifont}
\usepackage{hyperref}
\usepackage{multirow}
\usepackage[table]{xcolor}

\definecolor{stagegray}{gray}{0.94}

\setlist[itemize]{leftmargin=*,nosep}
\setlist[enumerate]{leftmargin=*,nosep}
\setlist[description]{leftmargin=*,nosep}

\newcommand{\tool}{\textsc{TATG}}

\newcommand{\rqanswer}[2]{%
  \vspace{0.35em}\noindent\fbox{%
  \begin{minipage}{0.94\linewidth}
  \textbf{Answer to RQ#1.} #2
  \end{minipage}}\vspace{0.55em}}

\definecolor{tatgblue}{RGB}{38,78,135}
\definecolor{tatgred}{RGB}{130,55,55}
\definecolor{listingbg}{RGB}{248,248,248}
\definecolor{listingframe}{RGB}{210,210,210}

\lstdefinestyle{tatgjava}{
  language=Java,
  basicstyle=\ttfamily\scriptsize,
  keywordstyle=\bfseries\color{tatgblue},
  commentstyle=\itshape\color{gray!70!black},
  stringstyle=\color{tatgred},
  numbers=left,
  numberstyle=\tiny\color{gray!70!black},
  stepnumber=1,
  numbersep=5pt,
  frame=single,
  framerule=0.3pt,
  rulecolor=\color{listingframe},
  backgroundcolor=\color{listingbg},
  breaklines=true,
  columns=fullflexible,
  keepspaces=true,
  showstringspaces=false,
  tabsize=2,
  captionpos=b,
  xleftmargin=1.6em,
  framexleftmargin=1.4em
}

\lstdefinestyle{tatgprompt}{
  basicstyle=\ttfamily\scriptsize,
  numbers=left,
  numberstyle=\tiny\color{gray!70!black},
  numbersep=5pt,
  frame=single,
  framerule=0.3pt,
  rulecolor=\color{listingframe},
  backgroundcolor=\color{listingbg},
  breaklines=true,
  columns=fullflexible,
  keepspaces=true,
  showstringspaces=false,
  tabsize=2,
  captionpos=b,
  xleftmargin=1.6em,
  framexleftmargin=1.4em
}

\newif\ifshowchanges
\showchangesfalse 
\newif\ifshowminorchanges
\showminorchangesfalse
\newcommand{\add}[1]{%
  \ifshowchanges
    {\color{blue}#1}%
  \else
    {#1}%
  \fi
}

\providecommand{\add}[1]{#1}
\renewcommand{\add}[1]{#1}

\makeatletter
\def\ps@headings{%
  \def\@oddhead{}%
  \def\@evenhead{}%
  \def\@oddfoot{\hfil\thepage\hfil}%
  \def\@evenfoot{\hfil\thepage\hfil}%
}
\makeatother

\makeatletter
\let\tatg@makecaption\lst@makecaption
\renewcommand{\lst@makecaption}[2]{%
  \setlength{\parindent}{0pt}%
  \addtolength{\hsize}{-\lst@xleftmargin}%
  \addtolength{\hsize}{-\lst@xrightmargin}%
  \tatg@makecaption{#1}{#2}%
}
\makeatother

\hypersetup{
  hidelinks,
  pdftitle={TATG: Tracking-Aware Testing Objective for LLM-based Test Generation},
  pdfauthor={Guancheng Wang, Qinghua Xu, Lionel Briand}
}

\begin{document}


\title{TATG: Tracking-Aware Testing Objective for LLM-based Test Generation}


\author{%
\begin{tabular}{@{}c@{}}
Guancheng Wang$^{*}$,
Qinghua Xu$^{*}$,
Lionel Briand$^{*\dagger}$\\[4pt]
{\normalsize $^{*}$ Research Ireland Lero Centre for Software,
University of Limerick, Limerick, Ireland}\\
{\normalsize $^{\dagger}$ School of Electrical Engineering and Computer Science,
University of Ottawa, Ottawa, Canada}\\
{\normalsize
guancheng.wang@ul.ie,
et.qinghua@gmail.com,
lionel.briand@lero.ie}
\end{tabular}%
}


\markboth{}{}


\IEEEtitleabstractindextext{%

\begin{abstract}

\add{Complex Java methods challenge automated unit test generation, as achieving high coverage and fault detection requires satisfying intricate testing requirements involving program states, execution paths, and assertions. Existing LLM-based approaches leverage program context, static analysis, and dynamic feedback but lack explicit tracking of individual testing requirements across iterations, potentially leading to redundant generation and unresolved testing objectives.

We present TATG, a tracking-aware LLM-based unit test generation framework that unifies structural coverage and mutation-based fault detection. TATG introduces a structured objective representation capturing testing requirements derived from static analysis and dynamic feedback, enabling fine-grained tracking of their satisfaction status. It employs a two-stage workflow that first improves structural coverage and subsequently strengthens assertions through mutation-guided hardening.

We evaluate TATG on 141 complex Java methods against state-of-the-art baselines, including KTester and PANTA. Results show average improvements of 22.15, 20.14, and 37.66 percentage points in line coverage, branch coverage, and mutation score, respectively. TATG also outperforms general-purpose coding assistants, including Codex and Claude Code, under the same open-weight model configurations, achieving higher average coverage and mutation scores while generating fewer tests and consuming fewer tokens. These findings highlight the value of explicitly tracking testing objectives for effective LLM-based test generation.}

\end{abstract}


}


\maketitle
\IEEEdisplaynontitleabstractindextext


\section{Introduction}
\label{sec:intro}

\IEEEPARstart{A}{utomated} unit test generation has relied on random testing, search-based testing, symbolic execution, and, more recently, large language models (LLMs). Despite this progress, generating high-quality tests for complex Java methods remains challenging. The difficulty often lies not in the size of the focal method itself, but in satisfying branch-specific testing requirements that are only partially visible from the method's source code. Reaching a target behavior may require establishing a prerequisite program state, exercising a particular switch branch, triggering an exception-translation path, invoking the method through a valid entry route, or constructing assertions strong enough to detect semantic regressions. Some of these requirements can be inferred statically, while others only become visible after generated tests are compiled, executed, and evaluated.

Recent LLM-based test generation approaches have moved beyond focal-method-only prompting by incorporating project context, static analysis, coverage feedback, and mutation guidance~\cite{wang2024hits,li2025ktester,gu2025llm,gu2024testart,altmayer2025coverup,liu2023pre,dakhel2024effective,harman2025mutation,wang2025mutation}. For example, KTester injects project structure and usage knowledge~\cite{li2025ktester}, PANTA uses coverage feedback to target under-tested paths~\cite{gu2025llm}, and MUTGEN leverages mutation feedback to strengthen generated tests~\cite{wang2025mutation}. However, these signals are typically delivered to the LLM as prompts, hints, path descriptions, or post-execution feedback. Existing approaches do not maintain an explicit representation of individual testing requirements and their status across iterations. Consequently, later rounds may revisit already satisfied behaviors while unresolved branches, or weak oracles remain unaddressed. Furthermore, coverage-oriented and mutation-oriented feedback are usually handled separately, limiting their ability to jointly improve reachability and fault-detection strength.

We present \tool{}, a framework that unifies static program analysis and dynamic evaluation feedback through a unified objective representation. \tool{} represents each testing requirement as a tracked objective containing fields such as supporting evidence, required setup, and target action. Static analysis generates structural objectives for reachability, branch coverage, exception handling, object construction, boundary behavior, and entry routes. During generation, compilation/execution, coverage, oracle-quality, and mutation feedback are converted into feedback objectives and integrated into the same tracking mechanism. This representation enables the framework to reason explicitly about which requirements have been satisfied, which remain unresolved, and which require further refinement.

Building on this objective representation, \tool{} employs a two-stage generation strategy. The structural stage focuses on producing executable tests, improving reachability, and increasing line/branch coverage using static objectives, along with compilation/execution and coverage feedback. Once structural progress stabilizes, the hardening stage uses mutation feedback to improve fault-detection capability. 

We evaluate \tool{} on an extended benchmark of 141 focal methods, comprising complex methods with rich control-flow structures and project dependencies, collected from all projects used by recent LLM-based baselines, as well as additional modern multi-module projects. We compare \tool{} against six representative approaches across three categories: traditional, hybrid (traditional + LLM), and LLM-based. For all approaches, we conduct the evaluation using two recently released open-source LLMs, Qwen3.6-35B-A3B and DeepSeek-V4-Flash-284B, representing different model sizes. We chose to rely on non-paid, local models as a more affordable alternative for most software development organizations and a way to minimize our experiment costs. 
The results show that \tool{} consistently outperforms all baselines in terms of structural coverage and mutation score under the same iteration budget. 


\add{We further evaluate \tool{} against two general-purpose coding assistants, Codex CLI and Claude Code, to investigate whether their agentic capabilities can match those of a task-specific test-generation framework. Although these assistants are commonly associated with proprietary LLMs, both support configurable model backends, allowing developers to use open-weight models~\cite{openai2026codexconfig,anthropic2026gateway}. This flexibility is particularly relevant in industrial settings, where reliance on externally hosted proprietary models may raise concerns regarding substantial API costs, usage restrictions, and data confidentiality and security~\cite{klemmer2024aiassistants,snyk2025aicode}. We therefore configure both assistants with the same two open-weight LLMs as \tool{}, enabling a controlled comparison of their test-generation strategies while reflecting practical deployment scenarios. Our results show that both assistants achieve lower test effectiveness than \tool{}. Moreover, unlike task-specific test-generation frameworks, these general-purpose coding assistants may extensively explore project repositories and incorporate additional context through autonomous tool interactions, making their context consumption less predictable and harder to control. Our experiments show that both assistants consume substantially more tokens than \tool{}, highlighting an important consideration in industrial environments where token usage and monetary budgets are often constrained~\cite{github2026copilotbudget,anthropic2026enterpriseusage}.}

\smallskip\noindent\textbf{Contributions.} This paper makes the following contributions:

\begin{itemize}

\item \textbf{Tracking-aware objective representation.}
We introduce a unified objective representation for LLM-based unit test generation that captures branch-level testing requirements together with supporting evidence, required setup, target actions, expected oracles, acceptance predicates, and execution status. The representation enables explicit tracking of which testing objectives have been satisfied and which remain unresolved across generation iterations.

\item \textbf{Two-stage objective-guided generation.}
We propose a two-stage objective-guided framework that coordinates structural coverage improvement and mutation-based oracle strengthening within a unified generation process. Static-analysis results, compilation/execution diagnostics, coverage feedback, and mutation feedback are all represented and managed through the same objective-tracking mechanism.

\item \add{\textbf{Comprehensive evaluation on challenging Java methods.}
We evaluate \tool{} on 141 complex Java methods, including the 110 KTester subjects and 31 additional challenging methods. We are not comparing LLMs, but alternative approaches to unit test generation, using the same recent open-weight models for all LLM-based approaches and selecting two models of very different sizes. Results show that \tool{} consistently outperforms strong state-of-the-art baselines and general-purpose coding assistants, including Codex and Claude Code, under the same open-weight model configurations, while generating fewer tests and consuming fewer tokens. Ablation studies further confirm the effectiveness of the two-stage generation strategy.}

\end{itemize}
\section{Motivating Example}
\label{sec:motivating}

We use a single running example to illustrate the testing problem that \tool{} addresses. Listing~\ref{lst:running} presents a simplified, representative token-stream reader that abstracts recurring testing challenges from real-world code, rather than reproducing a concrete class from an existing project.
The field \texttt{currToken} records the reader's current state.  The collaborator \texttt{stream} provides three operations: \texttt{advance()} moves to the next token and may throw \texttt{StreamException}; \texttt{getText()} returns the current string value; and \texttt{getInt()} returns the current integer value.  The method under test, \texttt{nextTextValue()}, returns a textual representation of the current token when possible and throws \texttt{IOException} when stream processing fails.  

\begin{lstlisting}[style=tatgjava,caption={Running example: stateful entry, finite dispatch, and exception translation.},label={lst:running}]
class StreamTokenReader {
  private Token currToken;
  private TokenStream stream;

  String nextTextValue() throws IOException {
    if (currToken == Token.FIELD_NAME) {      // G
      currToken = stream.advance();
      if (currToken == Token.STRING)          // P1
        return stream.getText();
      if (currToken == Token.NULL)            // P2
        return null;
      if (currToken == Token.INT)             // P3
        return String.valueOf(stream.getInt());
    }
    if (currToken == Token.NULL)              // P4
      return null;
    try {
      stream.advance();
    } catch (StreamException e) {             // P5
      throw new IOException("Stream error", e);
    }
    throw new IOException("Not a text value"); // P6
  }
}
\end{lstlisting}

A useful test suite for Listing~\ref{lst:running} must satisfy requirements that are not explicit.  First, most paths are \emph{state-dependent}.  Paths P1--P3 are reachable only when \texttt{currToken} is \texttt{FIELD\_NAME} before the call and \texttt{stream.advance()} then returns a particular next token.  A test that simply constructs a fresh reader and calls the target may reach only the fall-through exception.  Second, the number of nested branches is \emph{finite}: P1, P2, and P3 correspond to distinct token constants, and each needs a distinct input or mock behaviour.  Third, P5 requires \emph{exceptional predecessor behaviour}: the collaborator must throw \texttt{StreamException} at the right call site so that the method executes the catch-rethrow block.  Finally, if the method is package-private or is normally accessed through a facade, the test must also select an executable entry route.

These requirements also explain why coverage should precede mutation refinement.  If tests do not reach P2, P3, or P5, mutants inside those branches provide little actionable information: mutation analysis mainly reports that the mutated statements were not covered.  
In contrast, coverage feedback reveals unresolved coverage obligations and directs subsequent test generation toward the prerequisite state or collaborator behaviour needed to satisfy them.
Once those branches are reached, mutation feedback becomes useful: covering the \texttt{Token.STRING} branch is insufficient if the test does not verify the returned text. A mutant that changes \texttt{return stream.getText()} to \texttt{return ""} may not be killed. At that point, the useful objective is no longer reachability, but oracle strengthening, e.g., asserting the exact text returned by the stream.

Existing generators struggle with this example because the information is not represented at the right granularity. Search-based tools optimize generated call sequences and coverage feedback, but the required setup for this method is highly specific: the test must place the reader in a particular token state or configure the stream to fail when \texttt{advance()} is executed. LLM-based tools can reason about the source and often generate readable tests, yet a generic branch- or exception-oriented prompt does not translate P1--P6 into distinct testing goals with required setup and measurable acceptance criteria. KTester, a recent state-of-the-art approach, combines project knowledge for test generation, however, the context remains coarse-grained (e.g., it does not capture predicate-specific information such as condition variables) and is not maintained as a per-path objective that records whether each branch or exception path has been attempted, satisfied, or left open. As a result, a passing test suite may still leave deep paths uncovered, while subsequent repair or regeneration lacks an explicit objective that identifies the unresolved branch or exception behavior requiring attention, such as a specific token-dispatch branch or an exception-translation path.

\tool{} addresses this example through two-stage objective-guided generation. 
Before generation, static analysis translates branch-relevant facts into explicit objectives, such as ``exercise the branch where \texttt{currToken == Token.STRING}'', ``exercise the branch where \texttt{currToken == Token.NULL} after entering the \texttt{FIELD\_NAME} state'', and ``configure \texttt{stream.advance()} to throw \texttt{StreamException} and assert the resulting \texttt{IOException}''. 
The structural stage first uses compilation and execution diagnostics, together with coverage feedback, to improve target reachability and code coverage. 
After each round, \tool{} updates the objective state to record which objectives were satisfied, which failed compilation/execution, and which remain open. 
Only after sufficient structural progress does mutation feedback become actionable: mutants in unreached code mainly indicate missing reachability, whereas live mutants in reached code may indicate either insufficient test inputs or weak assertions. 
Therefore, the hardening stage converts mutation testing feedback into semantic objectives that strengthen oracles and kill live mutants. 
Both stages share the same objective schema, but the generation focus shifts from improving reachability to strengthening oracles.

\section{Approach}
\label{sec:approach}

\subsection{Overview}

\label{sec:approach-overview}

\tool{} is a two-stage, objective-guided framework for LLM-based unit test generation.
Its central abstraction is an \emph{objective state}: a set of testing goals that records what should be tested, what evidence supports each goal, how a test should reach it, how success will be checked, and whether previous attempts have made progress.
An objective may ask the model to reach a branch, construct a required state, repair an invalid setup, strengthen a weak assertion, or kill a live mutant.
Across iterations, \tool{} updates the status of each tracked objective, unifying compilation/execution diagnostics, coverage gaps, and mutation results into a single objective rather than handling them independently.

\begin{figure*}[t]
    \centering
    \includegraphics[width=0.96\textwidth]{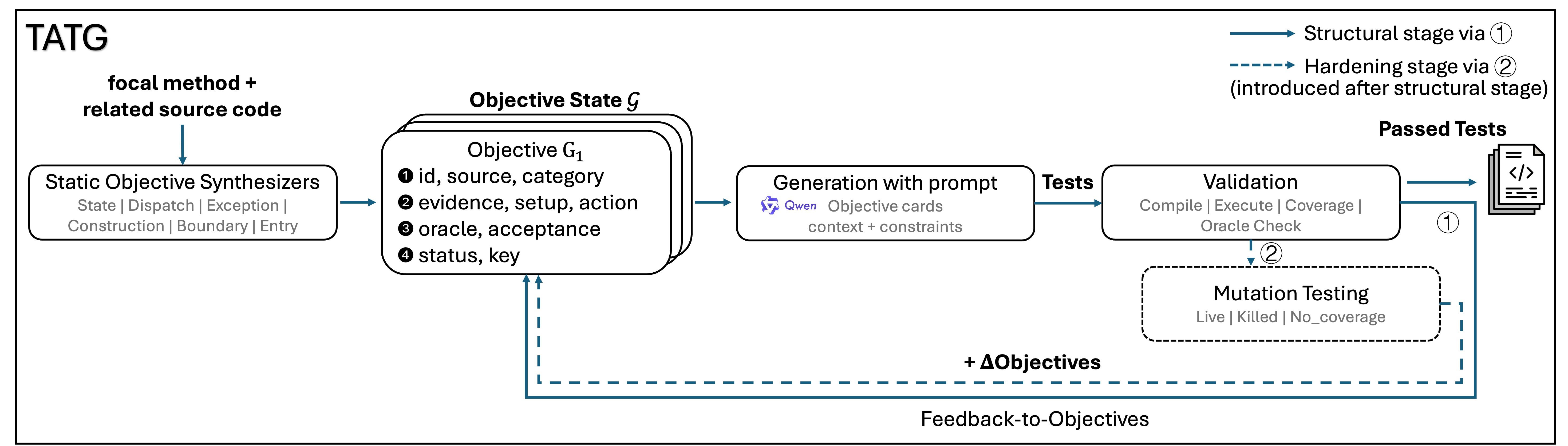}
    \caption{Overview of \tool{}.}
    \label{fig:overview}
\end{figure*}

Figure~\ref{fig:overview} shows the workflow.
Given a focal method and its corresponding source, \tool{} first constructs a compact target context that includes the method body, relevant condition values, construction information, and accessible entry routes.
Static objective synthesizers then create the initial objective state from source and type facts, such as state-dependent setup, branch-relevant facts, exception paths, construction constraints, boundary inputs, and entry routes.
The first stage performs structural generation, using compilation/execution diagnostics to produce executable tests and coverage feedback to improve branch reachability and coverage.
\add{The second stage performs hardening using feedback from a mutation testing tool---PITest in our experiments, as described in Section~\ref{sec:implementation}. This feedback is used to construct semantic objectives that guide the LLM to strengthen assertions and refine test inputs for mutants that are covered but not killed, while generating additional tests to exercise and kill previously uncovered mutants.}

The two stages share the same objective schema and the same objective-guided prompt structure, but they differ in focus.
The structural stage prioritizes executable tests and hard-to-reach paths.
The hardening stage prioritizes strengthening oracles for code that is already reachable.
This order is important for complex methods: if the test suite does not reach deep branches, mutation feedback often only reports that mutants are not covered, which provides little useful oracle guidance.

The remainder of this section follows the lifecycle of an objective in \tool{}.
Section~\ref{sec:objective-state} defines the objective;
Section~\ref{sec:obj-proj} describes how initial objectives are derived from the source code and how they are updated from compilation/execution diagnostics, coverage, and mutation feedback; 
Section~\ref{sec:prompt-construction} explains the prompt structure and how the objective state is rendered into the prompt;
and Section~\ref{sec:budgeting} describes the budget allocation and termination rules.

\subsection{Objective definition}
\label{sec:objective-state}

An objective state $\mathcal{G}$ is a set of objectives:
\[
\mathcal{G}=\{G_1,G_2,\ldots,G_n\},
\]
where each objective $G$ is represented as a tuple:
\[
G =
\langle
id, src, cat, E, S, A, Q, \alpha, status, key
\rangle .
\]
The tuple's fields have the following definitions.

\textbf{Identifier.}
The field \(id\) is an identifier for the objective, allowing \tool{} to track the same testing goal across iterations. 
It is derived from the objective source and its evidence. 
For example, an objective created from a missed branch may use the source line and branch index, while a mutation objective may use the mutant identifier reported by mutation testing. 

\textbf{Source.}
The field \(src\) records where the objective comes from:
\[
src \in
\{
static, coverage, validation, oracle, mutation
\}.
\]
Static objectives are created before generation. 
Coverage objectives are created from missed lines or branches. 
Validation objectives are created from compilation or execution failures. 
Oracle objectives are created when generated tests execute code, but contain weak or missing assertions. 
Mutation objectives are created from mutation testing feedback.

\textbf{Category.}
The field \(cat\) specifies the objective category, which determines how the objective is rendered in the prompt. 

\begin{description}[leftmargin=1.6em, style=nextline, font=\normalfont\itshape, itemsep=0.1em, topsep=0.2em]
\item[Structural categories.]
\textit{StateSetup} establishes object or collaborator state; 

\textit{DispatchArm} targets enum, switch, or constant-family branches; 

\textit{ExceptionPath} targets paths involving exception handling (e.g., try-catch and throw statements);

\textit{Construction} describes object or dependency construction; 

\textit{BoundaryInput} covers null, empty, malformed input, or boundary values; 

and \textit{EntryRoute} specifies how to reach non-directly callable targets, e.g., private focal methods.

\item[Feedback categories.]
\textit{ValidationRepair} repairs compilation, execution, or target-reach failures; 

\textit{OracleStrengthening} replaces weak observations with stronger assertions; 

\textit{MutationReachability} targets uncovered mutants; 

and \textit{MutationAssertion} adds assertions that distinguish the original behavior from a live mutant.
\end{description}

\textbf{Evidence.}
The field \(E\) contains the evidence that justifies the objective. 
Evidence may include a source line, a branch condition, an enum constant, a caught exception path, a constructor signature, a validation error, a weak assertion, or a live mutant description. 
For example, for a catch-rethrow path, \(E\) may contain the source snippet
\texttt{catch (StreamException e)} and the wrapper exception \texttt{IOException}. 
For a mutation objective, \(E\) may contain the mutator name, mutated line, and live mutant ID.

\textbf{Required setup.}
The field \(S\) describes what must be arranged before executing the target action. 
This may include constructing an object, selecting an input value, positioning an internal state through public APIs, configuring a mock collaborator, or routing through a facade method. 
For example, covering a \texttt{Token.NULL} branch under a \texttt{FIELD\_NAME} guard may require the setup ``advance the stream to \texttt{FIELD\_NAME}, then advance to \texttt{NULL}.''

\textbf{Target action.}
The field \(A\) describes the action that the test must execute. 
Usually, this is a direct call to the focal method. 
For non-public targets, \(A\) may be an accessible entry route, such as a public facade method that eventually reaches the target. 
For mutation objectives, \(A\) identifies the call intended to exercise the mutated code.

\textbf{Oracle shape.}
The field \(Q\) describes the expected form of the assertion, not necessarily the final assertion text. 
Oracle shapes include exact-value assertions, null assertions, exception assertions, state-change assertions, interaction assertions, and differential assertions. 
For example, an objective may specify that the test should use \texttt{assertThrows(IOException.class, ...)} for an exception path, or \texttt{assertEquals("42", result)} for a mutant that changes an integer-to-string return value.

\textbf{Acceptance predicate.}
The field \(\alpha\) is a measurable predicate that determines whether the objective has been satisfied after executing generated tests. 
It maps a feedback record to a Boolean value:
\[
\alpha(F) \rightarrow \{true,false\}.
\]
For a coverage objective, \(\alpha\) checks whether the corresponding line or branch is covered in the JaCoCo report. 
For a validation objective, \(\alpha\) checks whether the candidate compiles, executes, and reaches the target. 
For an oracle objective, \(\alpha\) checks whether the weak assertion has been replaced by a stronger oracle shape. 
For a mutation objective, \(\alpha\) checks whether a live mutant is killed.

\textbf{Status.}
The field \(status\) records the objective lifecycle:
\[
status \in \{open, attempted, satisfied, blocked\}.
\]
An objective is \emph{open} when it has not yet been addressed. 
It becomes \emph{attempted} when a generated test claims to target it. 
It becomes \emph{satisfied} when its acceptance predicate evaluates to true. 
It becomes \emph{blocked} when generated tests repeatedly fail validation or cannot reach the target action.

\textbf{De-duplication key.}
The field \(key\) records the behavioral signature of an objective attempt.
It is not a test name or a hash of the generated code.
Instead, it summarizes the behavior being attempted:
\[
key =
\langle
cat,\ S,\ A,\ Q
\rangle .
\]
For example, two tests that both exercise the \texttt{Token.NULL} arm after entering the \texttt{FIELD\_NAME} state and both assert a null result share the same key, even if their code differs syntactically.
When an objective repeatedly fails to make progress, either due to execution failures or inability to reach the target behavior, \tool{} retires the objective from $\mathcal{G}$ to avoid repeatedly exploring redundant generation strategies.

\begin{table*}[h]
\centering
\tiny
\caption{Projection from source-grounded evidence to objectives.}
\label{tab:static-objective-projection}
\begin{tabularx}{\textwidth}{
  p{2.1cm}
  >{\raggedright\arraybackslash}p{5.1cm}
  >{\raggedright\arraybackslash}X
  >{\raggedright\arraybackslash}X
  >{\raggedright\arraybackslash}X}
\toprule
Category \(cat\) & Evidence \(E\) from static analysis or dynamic feedback
& Required setup \(S\) & Target action \(A\) & Oracle shape \(Q\) \\
\midrule

\textit{StateSetup}
& Assignment, update, or predicate over a receiver field, e.g.,
\texttt{this.index++}, \texttt{\_token = p.nextToken()},
\texttt{if (this.index >= end)}, or a method call whose receiver resolves to a receiver
field, e.g., \texttt{this.context.inObject()}
& Establish the required receiver or collaborator state through visible constructors,
setters, builders, factories, or public methods
& Invoke the focal method after the required state has been established
& Assert return value, state transition, emitted output, collaborator interaction, or exception \\

\textit{DispatchArm}
& Branch or switch statement whose predicate/selector depends on a concrete
code element: \texttt{if (token == VALUE\_STRING)}, \texttt{switch (mode)},
\texttt{case STRICT:}, \texttt{"json".equals(format)}, \texttt{type == 1}, or
\texttt{x instanceof Foo}
& Choose argument values, receiver state, or concrete object types that satisfy the selected predicate, match the selected label, or satisfy the selected type test.
& Invoke the focal method, or an accessible caller that reaches the selected statement
& Assert behavior specific to that statement: return value, state, output, interaction,
or exception \\

\textit{ExceptionPath}
& Exception-handling paths involving \texttt{try}, \texttt{catch}, and optional
rethrowing, e.g., \texttt{try \{...\} catch (X e) \{throw new Y(e);\}}.
& Provide invalid input or configure a collaborator to raise the exception required by
the handler
& Invoke the focal method on the exceptional scenario
& Assert exception type, message, cause, wrapped exception, or documented fallback behavior \\

\textit{Construction}
& Object construction or initialization, including constructor calls, static factory methods, builder chains, concrete implementations of abstract/interface types, and field initialization visible in source (e.g., \texttt{this.client = client} in a constructor or setter)
& Build the receiver and arguments using visible construction APIs and available concrete implementations
& Invoke the focal method on the constructed object graph
& Assert concrete return value, state, output, interaction, or exception \\

\textit{BoundaryInput}
& Null, boundary, or shape predicate, including null checks, emptiness checks,
numeric or size comparisons, array/collection/string shape checks, and loop-bound conditions (e.g., \texttt{x == null}, \texttt{s.isEmpty()}, \texttt{list.size() > 1}, and \texttt{i < limit}).
& Instantiate values that satisfy or violate the target predicate under representative input conditions (e.g., null, boundary, validity, or shape)
& Invoke the focal method with the boundary-relevant input
& Assert exact value, nullness, exception, state change, or output \\

\textit{EntryRoute}
& Focal method has non-public visibility, or is reachable through a public or protected wrapper method
& Construct the receiver and arguments required by the accessible caller
& Invoke the shortest source-visible entry method that reaches the focal method
& Assert behavior observable through that entry method \\

\midrule

\textit{ValidationRepair}
& Compilation/execution diagnostics, or coverage reports
& Repair construction, imports, mocks, arguments, or setup so execution reaches the target branches
& Revise the test based on compilation/execution diagnostics or coverage feedback
& Preserve or add a concrete assertion after the repaired call \\

\textit{OracleStrengthening}
& Passing test with no assertion, only weak assertions (e.g., \texttt{assertNotNull}), or assertions that merely verify successful execution.
& Reuse the passing setup that already reaches the target behavior
& Execute the same target path
& Replace the weak check with exact-value, exception, state, interaction, output, or differential assertion \\

\textit{MutationReachability}
& Uncovered mutant reported on a source line that was not executed
& Arrange arguments or receiver state that execute the mutated statement
& Invoke the focal method so the mutated statement is reached
& Assert observable behavior after the mutated statement is reached \\

\textit{MutationAssertion}
& Live mutant on an already reached statement
& Arrange a mutant-reaching input whose behavior can differ from the original program
& Invoke the focal method killing the live mutant
& Assert exact or differential behavior that distinguishes the original program from the mutant \\

\bottomrule
\end{tabularx}
\end{table*}

\subsection{Objective Projection}
\label{sec:obj-proj}

Table~\ref{tab:static-objective-projection} summarizes how the objective fields (\textit{cat}, \(S\), \(A\), and \(Q\)) are instantiated from different types of evidence (\(E\)). We consider five evidence sources: static analysis, compilation/execution diagnostics, coverage feedback, oracle feedback, and mutation feedback. The first six rows correspond to structural objectives derived from source code, whereas the remaining rows correspond to feedback-driven objectives derived during iterative generation. For each evidence type, the table specifies how the setup (\(S\)), target action (\(A\)), and oracle shape (\(Q\)) are constructed.

\add{Structural objectives are initialized before the first generation round. Table I lists the complete set of objective categories used by TATG; these categories are shared by static analysis and LLM-based inference. Static analysis provides the initial objectives and supporting information for structural categories, including StateSetup, DispatchArm, ExceptionPath, Construction, BoundaryInput, and EntryRoute. To complement this initial extraction, TATG uses the objective definitions and the focal method source code to perform an LLM-based review of each category, identifying additional objective instances or relevant details that static analysis may have missed. 
For example, when examining Lines 17–22 of the running example, the LLM may identify a missing ExceptionPath objective for the try–catch–rethrow path and supplement a StateSetup requirement specifying the collaborator state needed to exercise that path. These complementary requirements guide test generation toward triggering catch block P5.}

Feedback-driven objectives are generated incrementally throughout iterative generation. Compilation and execution diagnostics, along with coverage reports, generate \textit{ValidationRepair} objectives; weak assertions identified in generated tests generate \textit{OracleStrengthening} objectives; and mutation testing generates \textit{MutationReachability} and \textit{MutationAssertion} objectives for uncovered and live mutants, respectively.

Table~\ref{tab:static-objective-projection} lists all structural objectives obtainable through static analysis. The additional LLM-based objective inference complements static analysis by identifying objectives that require semantic reasoning beyond what syntactic program analysis can provide. In contrast, feedback-driven objectives are introduced dynamically as new evidence becomes available after each generation round. Together, these objectives provide a unified representation that accumulates structural knowledge and execution feedback throughout the iterative generation process.

\subsection{Prompt Construction}
\label{sec:prompt-construction}

Listing~\ref{lst:prompt-structure} shows the prompt structure.
The central part of the prompt is the set of \emph{objective cards}. 
Each objective card is a textual rendering of one active objective. 
It contains the same fields as the objective state defined in Section~\ref{sec:objective-state}.
If the objective was updated in a previous round, the card also includes a brief prior observation explaining what happened, such as ``attempted but did not reach the branch'', ``compiled but failed at runtime'', or ``produced a live mutant''.

\begin{lstlisting}[style=tatgprompt,float=t,caption={Generation prompt structure used in each round.},label={lst:prompt-structure}]
Task:
  Generate or revise JUnit tests for the focal method.

Focal method context:
  focal_method:
    <method source>
  object_context:
    <relevant fields and helper methods>
  construction_context:
    <constructors, factories, builders, mocks, or subclasses>
  entry_context:
    <direct call or accessible caller routes>
  test_stack:
    <JUnit version, Mockito availability, Java version>

Active objective cards:
  Objective <G_i>:
    source:
      static | coverage | validation | oracle | mutation
    category:
      StateSetup | DispatchArm | ExceptionPath |
      Construction | BoundaryInput | EntryRoute |
      ValidationRepair | OracleStrengthening |
      MutationReachability | MutationAssertion
    status:
      open | attempted | blocked
    evidence:
      <source lines, branch condition, diagnostic,
       weak assertion, or mutant description>
    prior_observation:
      <what happened in the previous round, if any>
    required_setup:
      <object state, input values, mock behavior,
       construction strategy, or entry route>
    target_action:
      <direct target call or accessible caller route>
    oracle_shape:
      <exact value, null, exception, state change,
       interaction, or differential assertion>
    acceptance_condition:
      <how TATG will check whether G_i is satisfied>
    de_duplication_key:
      <cat, setup, action, oracle shape>

Global constraints:
  - compile under the project build
  - do not modify production code
  - use only available project dependencies
  - preserve already passing tests when repairing
  - avoid duplicating existing objective keys
  - return tests in the required format

Output contract:
  new_tests: [
    {
      objective_id: "<G_i>",
      test_name: "<name>",
      input_family: "<declared input partition>",
      setup: "<state/mocking/construction>",
      oracle_shape: "<declared oracle shape>",
      test_code_lines: ["<JUnit code lines>"]
    }
  ]
\end{lstlisting}

This structure allows each testing goal to receive its own feedback.
For example, if a previous test claimed to target the branch P2 in Listing~\ref{lst:running} but failed, \tool{} does not simply append ``missed branch at line 10'' to the next prompt. 
Instead, it updates the corresponding \textit{DispatchArm} objective: the evidence records the missed branch condition, the prior observation records the failed attempt, the required setup states how to reach the \texttt{NULL} state, the oracle shape specifies the expected assertion, and the acceptance condition states that P2 must be covered in the next run.

When a generated test fails to compile or execute, \tool{} creates or updates a \textit{ValidationRepair} objective. 
The evidence field records the compiler diagnostic, execution, or target-reach failure; the required setup field describes the suspected fix; the target action preserves the intended call or entry route; and the acceptance condition requires the revised test to compile, execute, and reach the target. 
Thus, repair is not handled by a separate prompt. 
It is represented as another objective category in the same objective state.

The LLM response fields are used to associate generated tests with objectives during validation. 
After execution, \tool{} updates each objective according to its acceptance condition and de-duplication key. 
This is the mechanism that turns multi-round prompting into explicit objective-state tracking.

\subsection{Budgeting and Termination}
\label{sec:budgeting}

\tool{} allocates the per-iteration generation budget according to the number of active objectives, rather than using a fixed number of tests for every round. 
The number of requested tests is computed as \(B_r=\mathrm{clip}(B_{\min}, B_{\max}, \lceil \lambda |\mathcal{A}_r| \rceil)\), where \(\lambda\) controls how many candidates are allocated per active objective, and \(B_{\min}\) and \(B_{\max}\) bound the budget. 
The concrete values of these parameters are reported in the evaluation setup. 
A stage terminates when its configured round limit is reached, when no active objective remains, or when recent rounds do not satisfy any new objectives. 



\section{Experimental Design}
\label{sec:eval}

We evaluate \tool{} by addressing the following research questions (RQs):

\begin{itemize}
\item \textbf{RQ1: Overall performance.}
How does \tool{} compare with representative baselines across effectiveness and efficiency metrics?

\item \textbf{RQ2: Comparison with coverage- and mutation-oriented approaches.}
How effective are the structural and hardening stages compared with representative coverage-oriented and mutation-oriented approaches, respectively?

\item \textbf{RQ3: Ablation study.}
How does each stage contribute to the overall effectiveness of \tool{}?

\item \add{\textbf{RQ4: Test generation with general-purpose coding assistants.}
How effective and efficient are Codex and Claude Code?}
\end{itemize}

RQ1 evaluates \ tool {} 's overall effectiveness and efficiency as a complete test-generation framework. RQ2 compares each stage with recent approaches targeting the same objective: structural coverage and mutation score, respectively. RQ3 isolates the contributions of the two stages through ablation analysis. Specifically, the structural stage produces executable tests and improves structural coverage, whereas the hardening stage uses mutation feedback to strengthen assertions and improve fault-detection capability. 

\add{RQ4 extends our evaluation to general-purpose coding assistants, namely Codex and Claude Code, to investigate whether their agentic capabilities can match those of a task-specific test generation framework. Both assistants support configurable LLM backends, enabling their use with open-weight models in addition to their default proprietary models. We therefore configure them with the same two open-weight LLMs as \tool{}, both to control for differences in underlying model capabilities and to reflect practical deployment scenarios where data confidentiality, API costs, and usage restrictions discourage reliance on proprietary models. We compare their test-generation effectiveness and resource consumption to assess the benefits of \tool{}’s task-specific design.}

We evaluate \tool{} on 141 complex Java methods, including all 110 subjects from the KTester benchmark and 31 additional challenging methods drawn from larger multi-module projects. The remainder of this section presents the datasets, baselines, evaluation metrics, and implementation.

\subsection{Datasets}

We adopt all focal methods from the KTester benchmark~\cite{li2025ktester}, which contains 110 complex Java methods from 10 open-source projects. To further evaluate \tool{} on even more challenging real-world software systems, we collect 31 additional methods from 10 open-source projects. Compared with the projects used by KTester, we select the additional projects from the MocklessTester benchmark~\cite{xu2026llm}. They include both Defects4J projects and modern multi-module Java projects (cutoff date: March 2025), such as JQuick-Curl~\cite{jquickcurl}, ADK Java~\cite{adkjava}, AgentScope Java~\cite{ascope}, and A2A Java~\cite{a2a}.
Given the high computational cost of evaluation, we carefully select these subjects through purposive sampling. We first prioritize projects not included in KTester to increase project diversity. Within these projects, we favor methods with relatively high cyclomatic complexity (CC) and non-comment lines of code (NLOC), as these characteristics generally indicate greater testing complexity. We further select methods that exhibit features known to challenge automated test generation, including package-private members (i.e., methods without an explicit access modifier), private methods, and dependencies that require constructing objects from external classes. Consequently, the additional subjects cover a broader range of access-control constraints and dependency patterns commonly encountered in real-world Java projects.

Table~\ref{tab:benchmark-projects} summarizes the benchmark subjects, including the number of methods, cyclomatic complexity (CC), and non-comment lines of code (NLOC). CC and NLOC are computed using Lizard~\cite{lizard}. Across all 141 methods, the mean CC is 18.5, the median CC is 16, and the mean NLOC is 59.8, indicating substantial structural complexity~\cite{ccpaper,wang2025mutation}. All methods are extracted from real-world open-source projects, ensuring that the evaluation is conducted on realistic testing tasks. \add{Compared with the KTester projects, the additional projects have substantially more production Java modules on average (10.0 vs.\ 1.2) and larger average production Java codebases (32.8 vs.\ 10.0 kNLOC), placing the selected methods in larger and more modular software contexts that may involve more complex dependencies and test setup, thereby posing additional challenges for test generation.}

\begin{table*}[t]
\centering
\caption{\add{Projects and focal method characteristics. CC and NLOC are computed by Lizard over the target Java source files.}}
\label{tab:benchmark-projects}
\tiny
\add{
\begin{tabular}{lllrrrrrrr}
\toprule
Project & Domain & Version & \#MUT & CC Mean & CC Median & CC Range & NLOC Mean & \shortstack{\#Java\\modules} & \shortstack{Project Java\\kNLOC} \\
\midrule
\multicolumn{10}{l}{\textbf{KTester benchmark methods (110)}} \\
Commons-CLI & Cmd-line Interface & 1.7.0-SNAPSHOT & 2
  & 12.0 & 12.0 & 10--14 & 36.5 & 1 & 2.3 \\
Commons-CSV & Data Processing & 1.10.0 & 6
  & 16.5 & 15.5 & 12--24 & 55.0 & 1 & 2.2 \\
Gson & Serialization & 2.10.1 & 20
  & 18.2 & 16.0 & 10--41 & 52.0 & 3 & 9.7 \\
Commons-codec & Encoding & 3a6873e & 18
  & 23.6 & 17.5 & 11--90 & 73.3 & 1 & 9.6 \\
Commons-collections4 & Utility & 4.5.0-M1 & 14
  & 17.4 & 16.0 & 10--29 & 54.5 & 1 & 30.7 \\
JDom2 & Text Processing (XML) & 2.0.6 & 21
  & 16.5 & 16.0 & 11--27 & 49.5 & 1 & 15.2 \\
Datafaker & Data Generation & 1.9.0 & 6
  & 15.2 & 14.0 & 9--25 & 39.0 & 1 & 12.0 \\
Event-ruler & Event Engine & 1.4.0 & 15
  & 21.5 & 15.0 & 11--45 & 68.5 & 1 & 6.2 \\
windward & Microservices & 1.5.1-SNAPSHOT & 2
  & 16.5 & 16.5 & 16--17 & 46.0 & 1 & 3.2 \\
batch-processing-gateway & Cloud Computing & 1.1 & 6
  & 16.0 & 17.5 & 10--19 & 67.0 & 1 & 8.6 \\
\midrule
Overall (110) & -- & -- & 110
  & 18.6 & 17.0 & 9--90 & 57.5 & 1.2 & 10.0\\
\midrule
\multicolumn{10}{l}{\textbf{Additional methods collected (31)}} \\
Commons-Math3 & Math / Optimization & Math-2f & 3
  & 32.1 & 31.0 & 20--45 & 118.3 & 1 & 83.8 \\
Jackson-Dataformat-XML & XML Serialization & JacksonXml-5f & 2
  & 14.0 & 14.0 & 8--20 & 61.5 & 1 & 4.9 \\
Commons-Lang3 & Core Java Utility & Lang-4f & 1
  & 55.0 & 55.0 & 55--55 & 132.0 & 1 & 21.7 \\
Jackson-Core & JSON Streaming & JacksonCore-26f & 1
  & 54.0 & 54.0 & 54--54 & 181.0 & 1 & 26.0 \\
Joda-Time & Date/Time & Time-13f & 1
  & 43.0 & 43.0 & 43--43 & 148.0 & 1 & 27.2 \\
Jsoup & HTML Processing & Jsoup-93f & 1
  & 3.0 & 3.0 & 3--3 & 11.0 & 1 & 12.0 \\
jquick-curl & Command Parsing & ee84ed6 & 3
  & 10.3 & 12.0 & 5--14 & 35.7 & 1 & 2.0 \\
ADK Java & AI Agent Framework & 5ee51fd & 10
  & 7.6 & 4.5 & 1--23 & 32.9 & 8 & 34.0 \\
AgentScope Java & AI Agent Framework & 28ef28a1 & 5
  & 27.0 & 16.0 & 6--60 & 110.4 & 38 & 85.7 \\
A2A Java & Agent Protocol SDK & 0b867d0d & 4
  & 11.5 & 6.5 & 1--32 & 42.8 & 46 & 30.3 \\
\midrule
Overall (31) & -- & -- & 31
  & 18.3 & 12.0 & 1--60 & 68.0 & 9.9 & 32.8\\
\midrule
Overall & -- & -- & 141
  & 18.5 & 16.0 & 1--90 & 59.8 & 5.6 & 21.4\\
\bottomrule
\end{tabular}}
\end{table*}

\subsection{Baselines} \label{sec:baselines}

We compare \tool{} against representative traditional, hybrid, and LLM-based test generation approaches.

\begin{itemize}
\item \textbf{Randoop}~\cite{pacheco2007randoop}: a feedback-directed random testing tool that generates unit tests through randomized sequences of method and constructor invocations.

\item \textbf{EvoSuite}~\cite{fraser2011evosuite}: a search-based test generation tool that optimizes generated test suites toward structural coverage objectives.
\item \textbf{EvoObj}~\cite{lin2021graph}: an extension of EvoSuite that improves object-oriented test generation by addressing object construction challenges.
\item \textbf{KTester}~\cite{li2025ktester}: a recent LLM-based test generation approach that injects project-structure and project-usage knowledge into the generation process.
\item \textbf{PANTA}~\cite{gu2025llm}: a recent LLM-based approach that combines static analysis and coverage-guided feedback to target under-tested execution paths.
\item \textbf{MUTGEN}~\cite{wang2025mutation}: a recent LLM-based approach that targets mutation score through mutation-testing feedback, used for comparison with the hardening stage.

\end{itemize}

\subsection{Evaluation Metrics}

\textbf{Effectiveness metrics.}
\tool{} aims to improve both the structural coverage and the fault-detection capability of generated tests. Following prior work~\cite{li2025ktester}, we report \textbf{line coverage} and \textbf{branch coverage} at the method level measured using JaCoCo~\cite{jacoco}. Line coverage measures the percentage of executable lines in the focal method covered by the generated tests, while branch coverage measures the percentage of conditional branches exercised.

To evaluate fault-detection capability, we use \textbf{mutation score} as the primary effectiveness metric. Mutation score measures the percentage of generated mutants detected by the generated test suite and is computed as:

\begin{equation}
\text{Mutation Score} =
\frac{\text{\# killed mutants}}
{\text{\# total mutants}}
\times 100
\label{eq:mutation_score}
\end{equation}

Following prior work~\cite{gu2025llm,li2025ktester,wang2025mutation}, PITest~\cite{pitest} produces mutants using its default mutator set. A mutant is considered \emph{killed} if at least one generated test distinguishes the behavior of the mutated program from that of the original program, typically through a failed assertion or an unexpected exception. A higher mutation score indicates stronger fault-detection capability.

\textbf{Efficiency metrics.}
We report the average number of generated test cases and the average end-to-end execution time per focal method. The former measures the number of test methods in the final generated test class, while the latter measures the average time required to complete the entire testing workflow for each approach on a focal method.



\subsection{Implementation} \label{sec:implementation}

All benchmark subjects are built using Maven. Following the official documentation of each tool, we integrate EvoSuite, Randoop, EvoObj, JaCoCo, and PITest into the Maven pipeline for automated test generation, coverage measurement, and mutation testing, respectively. We used a per-focal-method generation budget of 900 seconds and fixed the random seed to 3 for EvoSuite, Randoop, and EvoObj. \add{For these approaches requiring multiple random seeds, we report the average results across five independent runs.} Each of the 141 focal methods was evaluated independently. Generated tests were installed in isolated, copied workspaces and then evaluated using JaCoCo and PITest for method-level coverage and mutation scores. Among the baselines, as neither Randoop nor PANTA natively supports method-level generation, both are executed independently for each focal method, with PANTA additionally prompted to generate tests only for the target focal method.

\tool{} is implemented in Python. Static analysis uses SootUp~\cite{sootup_github}, including call-graph construction, type analysis, constructor discovery, branch analysis, exception analysis, and entry-route extraction.
As described in Section~\ref{sec:budgeting}, we set the maximum number of generated tests per focal method \(B_{\max}\) to 50, the minimum number \(B_{\min}\) to 1, and the factor $\lambda$ to 1. Static analysis is limited to 300 seconds per method, and the overall test-generation process to 1800 seconds. We use five iterations for each of the structural and hardening stages, for a total of 10 test-generation iterations. \add{We set an upper limit of 1,800 seconds per focal method for \tool{}, although the actual execution time is typically much shorter. For EvoSuite, Randoop, and EvoObj, our preliminary experiments showed that increasing the time budget beyond 900 seconds did not further improve their performance. Therefore, to reduce experimental costs, we retained a 900-second budget for these baselines. Notably, this budget is substantially more generous than their default settings and the time budgets commonly adopted in prior studies.}

We deploy Qwen3.6-35B-A3B locally using vLLM~\cite{vllm} with its default configuration and a temperature of 0.2. We use the same configuration across all experiments.
To assess the generality of the results obtained with \tool{}, we also evaluate it on the DeepSeek-V4-Flash-284B model served via Fireworks AI~\cite{fireworkAI}, using the same configuration. Both models were released around April 2026. For both models, we set the maximum output tokens per generation to 16,384 to accommodate complex test cases while bounding the computational overhead of individual requests.

To ensure a fair comparison, all approaches are executed using the same project-specific Java version and JUnit~5 framework. Coverage is measured with JaCoCo, and mutation testing is performed with PITest using its default mutator set. \add{We run KTester and PANTA for 10 iterations each, while MUTGEN is run for five iterations, consistent with the hardening stage. For LLM-based approaches, we conduct a single run per method, as our analysis aggregates results across multiple iterations and many methods, which mitigates the impact of randomness.}

We use the native Codex CLI (v0.144.5)~\cite{codex_cli,codex_custom_provider} and Claude Code (v2.1.311)~\cite{claude_code,claude_code_proxy} clients. Codex is configured with a custom model provider supporting its Responses' API, while Claude Code is connected to a Messages-compatible endpoint through its base URL and model settings. 

Both agents use the same Qwen and DeepSeek deployments described previously as their model backends. Consistent with \tool{}, each trial is limited to a maximum of ten generation rounds and a cumulative time budget of 1,800 seconds, encompassing agent execution, output extraction, external compilation and testing, and coverage and mutation-testing feedback. 

We use both agents in a manner that closely resembles real-world development scenarios. Specifically, we provide each agent with the project path and the target method under test, and instruct it to generate a test suite that achieves high code coverage and fault-detection effectiveness. An external harness writes the generated test source code verbatim, without any modifications, and compiles and executes the tests. For successfully executed test suites, the harness collects focal-method coverage using JaCoCo and mutation-testing results using PITest. 

In subsequent rounds, the agents receive feedback on compilation errors, test failures, the number of generated tests, focal-method coverage, and mutation scores, allowing them to iteratively improve their test suites. The same native agent session is maintained across all rounds, preserving the interaction history and enabling the agents to refine their tests based on previous feedback. This setup reflects a practical agent-assisted testing workflow, in which developers specify testing objectives and agents iteratively generate and improve tests based on execution feedback.

All experiments are conducted on a Linux server running Ubuntu 24.04.1 LTS with a 56-core CPU, 125 GB of RAM, and two NVIDIA RTX 6000 Ada GPUs.
The implementation of \tool{} and the benchmark subjects will be made publicly available.
\section{Experimental Results}
\label{sec:evalresults}
\begin{table*}[t]
\centering
\caption{Overall coverage, mutation score, and cost metrics on 141 methods. Gray rows denote intermediate stage results included for subsequent stage-level analysis.}
\label{tab:eval-overall}
\scriptsize
\setlength{\tabcolsep}{4pt}
\begin{tabular}{lllrrrrr}
\toprule
LLM & Tool & Stage & Line (\%) & Branch (\%) & Mutation score (\%) & Avg. time (s) & Avg. tests \\
\midrule
\multirow{2}{*}{--}
  & EvoSuite & -- & 11.35 & 9.97  & 7.08  & 900    & 2.45 \\
  & Randoop  & -- & 21.52 & 17.17 & 14.31 & 900    & 8771.90 \\
  & EvoObj  & -- & 38.84 & 34.48 & 2.76  & 900    & 4.04 \\
\midrule

\multirow{8}{*}{Qwen3.6-35B-A3B}
  & KTester & -- & 44.26 & 40.02 & 28.92 & 554.16 & 15.78 \\
  & PANTA   & -- & 51.17 & 46.70 & 20.11 & 441.89 & 36.02 \\
  & \multirow{5}{*}{TATG}
  & \cellcolor{stagegray}Structural
  & \cellcolor{stagegray}62.65
  & \cellcolor{stagegray}51.67
  & \cellcolor{stagegray}40.54
  & \cellcolor{stagegray}355.20
  & \cellcolor{stagegray}8.10 \\
  &
  & \cellcolor{stagegray}$\Delta$ Hardening (+MUTGEN)
  & \cellcolor{stagegray}+11.00
  & \cellcolor{stagegray}+12.75
  & \cellcolor{stagegray}+21.55
  & \cellcolor{stagegray}+420.20
  & \cellcolor{stagegray}+3.44 \\
  &
  & \cellcolor{stagegray}Final (+MUTGEN)
  & \cellcolor{stagegray}73.65
  & \cellcolor{stagegray}64.42
  & \cellcolor{stagegray}62.09
  & \cellcolor{stagegray}775.40
  & \cellcolor{stagegray}11.54 \\
  &
  & \cellcolor{stagegray}$\Delta$ Hardening (+Default)
  & \cellcolor{stagegray}+12.63
  & \cellcolor{stagegray}+16.06
  & \cellcolor{stagegray}+24.43
  & \cellcolor{stagegray}+482.95
  & \cellcolor{stagegray}+2.00 \\
  &
  & \textbf{Final (+Default)}
  & \textbf{75.28}
  & \textbf{67.73}
  & \textbf{64.97}
  & 838.15
  & 10.10 \\

\midrule

\multirow{8}{*}{DeepSeek-V4-Flash-284B}
  & KTester & -- & 48.10 & 43.82 & 31.09 & 664.52 & 21.46 \\
  & PANTA   & -- & 59.72 & 51.97 & 34.46 & 523.66 & 42.87 \\
  & \multirow{5}{*}{TATG}
  & \cellcolor{stagegray}Structural
  & \cellcolor{stagegray}65.80
  & \cellcolor{stagegray}57.69
  & \cellcolor{stagegray}52.89
  & \cellcolor{stagegray}278.40
  & \cellcolor{stagegray}7.40 \\
  &
  & \cellcolor{stagegray}$\Delta$ Hardening (+MUTGEN)
  & \cellcolor{stagegray}+9.19
  & \cellcolor{stagegray}+8.43
  & \cellcolor{stagegray}+9.10
  & \cellcolor{stagegray}+390.14
  & \cellcolor{stagegray}+14.14 \\
  &
  & \cellcolor{stagegray}Final (+MUTGEN)
  & \cellcolor{stagegray}74.99
  & \cellcolor{stagegray}66.12
  & \cellcolor{stagegray}61.99
  & \cellcolor{stagegray}668.54
  & \cellcolor{stagegray}21.54 \\
  &
  & \cellcolor{stagegray}$\Delta$ Hardening (+Default)
  & \cellcolor{stagegray}+14.13
  & \cellcolor{stagegray}+13.52
  & \cellcolor{stagegray}+12.03
  & \cellcolor{stagegray}+355.80
  & \cellcolor{stagegray}+5.40 \\
  &
  & \textbf{Final (+Default)}
  & \textbf{79.93}
  & \textbf{71.21}
  & \textbf{64.92}
  & 634.20
  & 12.80 \\

\bottomrule
\end{tabular}
\end{table*}

\subsection{RQ1: Overall Performance}
\label{sec:eval-rq0}
We compare the end-to-end performance of the two-stage \tool{} with the first five baselines listed in Section~\ref{sec:baselines}. 
All LLM-based and hybrid approaches are evaluated using two recently released open-source LLMs, Qwen3.6-35B-A3B (35B) and DeepSeek-V4-Flash (284B), which represent two different model scales and have demonstrated strong code-generation performance. Our approach is the first LLM-based test generation approach that jointly targets structural coverage and mutation score. Since no existing LLM-based approach optimizes both objectives simultaneously, the end-to-end comparison is limited to recent coverage-oriented approaches: KTester and PANTA.

The complete results are reported in Table~\ref{tab:eval-overall}. \add{The unshaded rows report the end-to-end results used for the overall comparison in RQ1, with \textit{Final (+Default)} representing the complete two-stage \tool{} framework.}
Across both models, \tool{} consistently achieves the best overall performance in terms of line coverage, branch coverage, and mutation score while generating fewer test cases than the baselines. Specifically, \tool{} improves both structural coverage and fault-detection capability over existing approaches on both models, demonstrating that the proposed two-stage objective-tracking framework is effective regardless of the underlying LLM.
We further conducted paired-sample Wilcoxon signed-rank tests on line coverage, branch coverage, and mutation score. The results show that \tool{} achieves statistically significant improvements over both PANTA and KTester across all three metrics on both LLMs ($\alpha = 0.05$).

In terms of generation time, the primary overhead comes from invoking PITest during each hardening iteration to obtain mutation feedback. Although mutation testing accounts for most of the additional runtime, this overhead is acceptable in practice, as the hardening stage further improves mutation scores and structural coverage while increasing the average execution time by at most approximately 8 minutes across the two models, as shown by the $\Delta$ Hardening (+Default) results in Table~\ref{tab:eval-overall}.

For traditional approaches, EvoSuite and Randoop achieve nearly complete coverage on relatively easy subjects. However, their overall performance is substantially worse than alternatives because they fail to generate valid tests for many of the more challenging subjects, especially those with rich project dependencies. Although EvoObj enhances EvoSuite through graph-based seed object synthesis, its overall performance remains limited by the underlying search-based test generation framework when handling complex project dependencies.

\rqanswer{1}{Across both LLMs, the complete two-stage framework consistently achieves the best overall performance, improving line coverage, branch coverage, and mutation score by an average of 22.15, 20.14, and 37.66 percentage points, respectively, over the strongest baseline PANTA.}

\subsection{RQ2: Comparison with Coverage- and Mutation-Oriented Approaches}
\label{sec:eval-rq1}
\add{To evaluate the two-stage design of \tool{}, we compare its stages individually with approaches targeting the same objective. In the gray-shaded rows of Table~\ref{tab:eval-overall}, \textit{Structural} reports the results of the coverage-oriented stage, while $\Delta$ \textit{Hardening} reports the incremental changes introduced by mutation-oriented hardening relative to \textit{Structural}. The parenthesized labels specify the hardening approach: \textit{+MUTGEN} uses MUTGEN~\cite{wang2025mutation}, a recent approach targeting mutation score, whereas \textit{+Default} uses our proposed hardening method.}

Across both LLMs, the structural stage consistently outperforms all coverage-oriented baselines. Specifically, compared with the best baseline PANTA, as shown in Table~\ref{tab:eval-overall}, the structural stage of \tool{} improves line coverage by 11.48\% and branch coverage by 4.97\% on Qwen, and by 6.06\% and 5.72\% on DeepSeek. Notably, these improvements are achieved using only five iterations, whereas PANTA is configured with ten iterations. For the hardening stage, \tool{} consistently achieves higher mutation scores than MUTGEN, improving the mutation score by 2.88 \% on Qwen and by 2.93 \% on DeepSeek. Moreover, the hardening stage further improves the line and branch coverage by 13.38\% and 14.79\% on average, respectively, compared with 10.10\% and 10.59\% achieved by MUTGEN.

Both stages improve because of the unified objective representation. Instead of repeatedly prompting the LLM with newly obtained feedback at each iteration, \tool{} maintains explicit objective states throughout iterative generation. Completed objectives are not revisited, allowing subsequent iterations to focus exclusively on unresolved testing goals. Furthermore, each objective provides fine-grained guidance beyond less informative coverage or mutation feedback, including input constraints, dependency construction, and mutation-specific requirements. 
This objective-tracking mechanism enables the LLM to use each generation iteration more effectively, improving both structural coverage and mutation score.

\rqanswer{2}{Both stages outperform approaches targeting the same objective. The unified objective representation enables more effective iterative guidance by tracking unresolved testing objectives and providing fine-grained feedback.}

\subsection{RQ3: Ablation Study}
\label{sec:eval-rq2}
\add{To assess the contribution of the hardening stage, we compare the \textit{Structural} and \textit{Final (+Default)} rows in Table~\ref{tab:eval-overall}. The corresponding $\Delta$ \textit{Hardening (+Default)} rows quantify the changes introduced by our hardening stage, showing consistent improvements in mutation score over the structural stage alone across both LLMs.}

To evaluate the structural stage's contribution, we conduct an ablation study by removing it and running only the hardening stage for 10 iterations. Since PITest requires at least one executable test to produce mutation feedback, the first iteration is performed without mutation feedback to bootstrap test generation. Without the structural stage, \tool{} achieves only 26.90\% line coverage, 20.45\% branch coverage, and a mutation score of 23.45\% on Qwen. The corresponding results on DeepSeek are 24.38\%, 18.37\%, and 20.91\%, respectively. Performance drops substantially in both structural coverage and mutation score. This result supports our design rationale that effective mutation-guided hardening first requires sufficient reachability and structural coverage.

\add{We did not conduct an ablation study removing objective tracking, as it is a fundamental component of \tool{} that supports the effective operation of both stages. Moreover, prior ablation studies by KTester~\cite{li2025ktester} and PANTA~\cite{gu2025llm} have shown the limitations of relying solely on basic method information for test generation. For instance, PANTA reported reductions of more than 50\% in both line and branch coverage when only basic method information was retained. Since \tool{} incorporates both basic method information and objective-tracking information into its prompts, removing objective tracking would leave the framework with substantially less guidance for test generation. Based on these findings, we expect such an ablation to result in substantial performance degradation across evaluation metrics. We therefore prioritized evaluating other design components rather than ablating this foundational mechanism.}

\rqanswer{3}{The ablation study demonstrates that each stage of \tool{} contributes to its overall performance. In particular, the structural stage plays a critical role by establishing stronger structural coverage, thereby improving the reachability of subsequent mutation-based hardening.}

\begin{table*}[t]
\centering
\caption{\add{Coverage, mutation score, and cost of general-purpose coding assistants on 141 methods.}}
\label{tab:eval-agents}
\scriptsize
\setlength{\tabcolsep}{4pt}
\add{%
\begin{tabular}{llrrrrr}
\toprule
LLM & Tool & Line (\%) & Branch (\%) & Mutation score (\%) &
Avg. time (s) & Avg. tests \\
\midrule
\multirow{2}{*}{Qwen3.6-35B-A3B}
  & Codex       & 40.00 & 39.00 & 37.44 & 638.21 & 56.60 \\
  & Claude Code & 51.51 & 51.64 & 49.66 & 692.40 & 66.75 \\
\midrule
\multirow{2}{*}{DeepSeek-V4-Flash-284B}
  & Codex       & 58.05 & 58.52 & 56.27 & 584.61 & 34.95 \\
  & Claude Code & 69.46 & 68.44 & 62.21 & 575.72 & 39.30 \\
\bottomrule
\end{tabular}}
\end{table*}

\subsection{RQ4: Test Generation with General-Purpose Coding Assistants}
\label{sec:eval-rq4}

\add{We evaluate Codex and Claude Code on the same 141 focal methods using two model backends, Qwen and DeepSeek, consistent with the experimental configurations used for \tool{}. As discussed in Section~\ref{sec:intro}, our focus on open-weight models reflects practical deployment considerations, particularly for developers and organizations seeking to manage API costs, usage restrictions, and data confidentiality and security risks associated with externally hosted models. Both coding assistants support configurable model backends, allowing them to operate with open-weight LLMs rather than their default proprietary models. By configuring all approaches with the same underlying LLMs, we aim to compare their test-generation strategies and agentic frameworks while controlling for differences in model selection.  Table~\ref{tab:eval-agents} presents the results under the evaluation protocol described in Section~\ref{sec:implementation}.}

\add{With Qwen, Claude Code outperforms Codex across all three effectiveness metrics, achieving higher effective line coverage (51.51\% vs.\ 40.00\%), branch coverage (51.64\% vs.\ 39.00\%), and mutation score (49.66\% vs.\ 37.44\%). Claude Code also generates more test entries on average (66.75 vs.\ 56.60), although it requires slightly more time for generation and feedback (692.40 vs.\ 638.21 seconds).}

\add{With DeepSeek, Claude Code consistently achieves higher effectiveness than Codex, with effective line coverage of 69.46\% vs.\ 58.05\%, branch coverage of 68.44\% vs.\ 58.52\%, and mutation scores of 62.21\% vs.\ 56.27\%. Unlike the Qwen configuration, Claude Code also requires slightly less time (575.72 vs.\ 584.61 seconds), while generating more test entries on average (39.30 vs.\ 34.95). These results suggest that the choice of coding assistant can substantially influence test-generation effectiveness, even when the underlying model backend is held constant.}

\add{We further conducted paired Wilcoxon signed-rank tests to compare \tool{} with the four coding assistant configurations in terms of effective line coverage, branch coverage, and mutation score. The results show that \tool{} achieves statistically significant improvements across all three metrics ($\alpha = 0.05$), except for branch coverage and mutation score when compared with Claude Code using DeepSeek, where the differences are not statistically significant. Nevertheless, \tool{} achieves higher average values for both metrics while generating substantially fewer tests (12.80 vs.\ 39.30 per method). Claude Code with DeepSeek requires slightly less generation time (575.72 vs.\ 634.20 seconds per method). However, this comparison involves different execution frameworks: Claude Code is a mature, actively maintained general-purpose coding assistant, whereas \tool{} employs a research-oriented, task-specific generation harness. The difference in execution time should therefore be interpreted in light of these implementation differences rather than as evidence of an inherent efficiency advantage.}

\add{The results also reveal the considerable resource demands of applying general-purpose coding assistants to focal-method test generation. With DeepSeek, Codex and Claude Code consume an average of 1,586,694 and 1,786,749 tokens per method, respectively, compared with 759,595 tokens for \tool{}. Similarly, with Qwen, Codex and Claude Code consume 1,599,44 and 1,777,412 tokens per method, respectively. In contrast, \tool{} explicitly incorporates static-analysis-derived testing objectives and task-specific context to guide test generation. Such targeted guidance reduces unnecessary project exploration and focuses generation on relevant program behaviors.}

\rqanswer{4}{\add{\tool{} consistently outperforms general-purpose coding assistants, including Codex and Claude Code, in test-generation effectiveness under the same open-weight model configurations, achieving significantly higher average coverage and mutation scores while generating much fewer tests and consuming fewer tokens. 
These findings highlight the effectiveness and efficiency of \tool{} as a task-specific test generation framework.
}}

\subsection{Threats to Validity}
\label{sec:threats}

\textbf{Internal validity.}
LLM outputs are stochastic and may vary across model releases, inference settings, and execution environments. 
To reduce this threat, we evaluate all LLM-based approaches with the same model version, configuration, prompting budget, and execution environment, and apply the same validation, coverage, and mutation testing pipeline to all generated tests.

\textbf{Construct validity.}
Construct validity may be affected by whether the selected metrics accurately reflect the objectives of the proposed approach. Since \tool{} explicitly targets structural coverage and mutation score, we evaluate it using method-level line coverage, branch coverage, and mutation score, following prior work.

\textbf{External validity.}
External validity is limited by the benchmark and implementation scope. 
We reuse benchmark tasks, project versions, and evaluation protocols from prior work~\cite{li2025ktester} and add 31 challenging methods from larger multi-module projects to enable broader assessment. 
The current implementation targets Java/JUnit. 
While \tool{}'s objective representation is language-independent, porting \tool{} to another ecosystem would simply require replacing the Java-specific providers used for static analysis, test execution, coverage measurement, mutation testing, and validation diagnostics, such as SootUp, JaCoCo, and PITest.
\add{Furthermore, our evaluation considers two recent, state-of-the-art open-weight LLMs, which may limit the generalizability of our findings to other models, particularly proprietary LLMs.
Although using the same underlying models enables controlled comparisons between \tool{} and general-purpose coding assistants, their relative effectiveness may differ across model families and capability levels.
Evaluating with alternative approaches with proprietary LLMs presents practical challenges, including substantial API costs, usage quotas, and potential data confidentiality and intellectual property risks associated with transmitting source code to external providers~\cite{nist2024genai,owasp2025llm}.
Moreover, provider-controlled model updates and service availability may affect experimental reproducibility~\cite{sallou2024breaking}.
Extending our evaluation to additional programming languages and LLMs, including proprietary models, remains an important direction for future work.}
\section{Related Work}
\label{sec:related}

We organize existing test generation approaches by the type of guidance used during the generation process. Recent approaches increasingly integrate multiple sources of guidance, including static analysis, coverage, and mutation feedback, to iteratively refine generated tests. The key distinction between existing work and \tool{} lies in how these guidance signals are represented, managed, and propagated across iterations.

\paragraph{Traditional automated test generation}
Search-based, random, and symbolic techniques predate LLMs and demonstrate the power of explicit coverage objectives and execution feedback.  Tools such as EvoSuite~\cite{fraser2011evosuite} and its many‐objective extensions (e.g., MOSA~\cite{suppapitnarm2000simulated}, DynaMOSA~\cite{panichella2017automated}) optimise towards structural coverage targets, while Randoop~\cite{panichella2017automated} uses feedback-directed random exploration; symbolic executors like KLEE~\cite{cadar2008klee} and JPF~\cite{visser2004test} solve path constraints to synthesise inputs. These techniques have strong coverage and fault-detection potential, but they rely primarily on search heuristics, branch-distance functions, and path constraints rather than explicit representations of testing objectives, making it difficult to capture the complex project dependencies required to generate executable tests.

\paragraph{LLM-based and knowledge-guided generation}
The emergence of large language models has prompted recent techniques that treat test generation as a code understanding and synthesis task.  
Approaches such as \emph{AthenaTest, TestSpark, ASTER, HITS, RAGTest, ChatUniTest, TestSpark} and \emph{KTester} guide LLMs by incorporating additional information into prompts through program analysis or retrieval techniques~\cite{tufano2020unit,sapozhnikov2024testspark,pan2025aster,wang2024hits,shin2024retrieval,chen2024chatunitest,sapozhnikov2024testspark}.
These approaches highlight the importance of contextual guidance.  However, testing intent and supporting knowledge are typically embedded in prompts ad hoc, making it difficult to represent, organize, and systematically reuse testing objectives.

\paragraph{Feedback‑guided improvement}
Several approaches iteratively refine tests using execution or coverage feedback~\cite{alshahwan2024automated,altmayer2025coverup,liu2023pre,lemieux2023codamosa,gu2024testart}.  
Among them, \emph{PANTA} alternates static control‑flow analysis with dynamic coverage measurement to drive an iterative generate–execute loop and systematically explore uncovered branches~\cite{gu2025llm}.
These approaches underscore the value of feedback loops but primarily treat feedback as prompt text, coverage summaries, or repair requests without maintaining a structured state across iterations.  In contrast, \tool{} normalises static and dynamic signals into a unified objective representation. Feedback thus becomes explicit and traceable, enabling unified management of structural and feedback-driven objectives across iterations.

\paragraph{Mutation‑guided generation}
Mutation testing provides a stronger adequacy signal than structural coverage~\cite{wang2025mutation,pitest}.  Recent approaches, such as \emph{MuTAP}~\cite{dakhel2024effective}, \emph{ACH}~\cite{alshahwan2024automated}, and \emph{MUTGEN}~\cite{wang2025mutation}, integrate mutation feedback into LLM‑based generation to strengthen assertions.  
These approaches demonstrate the effectiveness of mutation signals but treat mutants as separate targets.  \tool{} includes mutation objectives in the same tracking model as structural and feedback-driven objectives; once reachability and basic assertions are achieved, mutation objectives trigger oracle strengthening within the two-stage generation process.


\section{Conclusion}
\label{sec:conclusion}

This paper presents \tool{}, an LLM-based unit test generation framework based on a unified objective representation of test objectives. By representing static analysis results, execution feedback, coverage gaps, and mutation feedback as tracked objectives, \tool{} enables iterative guidance across generation rounds, allowing the LLM to focus on unresolved testing objectives.

\add{We evaluated \tool{} on 141 complex Java methods against representative traditional, hybrid, and LLM-based test generation approaches. For the latter, we compare generation approaches with identical open-weight configurations, as explained in the introduction. 

Experimental results show that \tool{} consistently achieves higher structural coverage and mutation scores than existing baselines. Compared with general-purpose coding assistants, including Codex and Claude Code, \tool{} also achieves higher average coverage and mutation scores under the same open-weight model configurations. 

Overall, our empirical findings highlight the effectiveness of unified objective tracking in guiding iterative test generation and improving both structural coverage and fault-detection capability.}


\section*{Acknowledgment}
This work has emanated from research jointly funded by Taighde Éireann -- Research Ireland under Grant number~13/RC/2094\_2 and by Huawei Technologies Co., Ltd. Lionel Briand is also supported by the Natural Sciences and Engineering Research Council of Canada. For the purpose of Open Access, the authors have applied a CC BY public copyright licence to any Author Accepted Manuscript version arising from this submission.


\bibliographystyle{IEEEtran}
\bibliography{references}

@article{gu2025llm,
  title={LLM Test Generation via Iterative Hybrid Program Analysis},
  author={Gu, Sijia and Nashid, Noor and Mesbah, Ali},
  journal={arXiv preprint arXiv:2503.13580},
  year={2025}
}

@inproceedings{fraser2011evosuite,
  title={Evosuite: automatic test suite generation for object-oriented software},
  author={Fraser, Gordon and Arcuri, Andrea},
  booktitle={Proceedings of the 19th ACM SIGSOFT symposium and the 13th European conference on Foundations of software engineering},
  pages={416--419},
  year={2011}
}

@inproceedings{pacheco2007randoop,
  title={Randoop: feedback-directed random testing for Java},
  author={Pacheco, Carlos and Ernst, Michael D},
  booktitle={Companion to the 22nd ACM SIGPLAN conference on Object-oriented programming systems and applications companion},
  pages={815--816},
  year={2007}
}

@inproceedings{cadar2008klee,
  title={Klee: unassisted and automatic generation of high-coverage tests for complex systems programs.},
  author={Cadar, Cristian and Dunbar, Daniel and Engler, Dawson R and others},
  booktitle={OSDI},
  volume={8},
  pages={209--224},
  year={2008}
}

@misc{sootup_github,
  author       = {{soot-oss}},
  title        = {SootUp: A Redesign of the Soot Static Analysis Framework},
  year         = {2023},
  howpublished = {\url{https://github.com/soot-oss/SootUp}},
  note         = {Accessed: 2026-02-05}
}

@article{wang2025mutation,
  title={Mutation-Guided Unit Test Generation with a Large Language Model},
  author={Wang, Guancheng and Xu, Qinghua and Briand, Lionel C and Liu, Kui},
  journal={arXiv preprint arXiv:2506.02954},
  year={2025}
}

@inproceedings{wang2024hits,
  title={Hits: High-coverage llm-based unit test generation via method slicing},
  author={Wang, Zejun and Liu, Kaibo and Li, Ge and Jin, Zhi},
  booktitle={Proceedings of the 39th IEEE/ACM International Conference on Automated Software Engineering},
  pages={1258--1268},
  year={2024}
}

@inproceedings{pan2025aster,
  title={Aster: Natural and multi-language unit test generation with llms},
  author={Pan, Rangeet and Kim, Myeongsoo and Krishna, Rahul and Pavuluri, Raju and Sinha, Saurabh},
  booktitle={2025 IEEE/ACM 47th International Conference on Software Engineering: Software Engineering in Practice (ICSE-SEIP)},
  pages={413--424},
  year={2025},
  organization={IEEE}
}

@article{liu2023pre,
  title={Pre-train, prompt, and predict: A systematic survey of prompting methods in natural language processing},
  author={Liu, Pengfei and Yuan, Weizhe and Fu, Jinlan and Jiang, Zhengbao and Hayashi, Hiroaki and Neubig, Graham},
  journal={ACM computing surveys},
  volume={55},
  number={9},
  pages={1--35},
  year={2023},
  publisher={ACM New York, NY}
}

@article{li2025ktester,
  title={KTester: Leveraging Domain and Testing Knowledge for More Effective LLM-based Test Generation},
  author={Li, Anji and Liu, Mingwei and Chen, Zhenxi and Pei, Zheng and Li, Zike and Dai, Dekun and Wang, Yanlin and Zheng, Zibin},
  journal={arXiv preprint arXiv:2511.14224},
  year={2025}
}

@misc{jacoco,
title = "JaCoCo",
year = "Accessed: 2026",
url = {https://www.jacoco.org}
}

@misc{pitest,
title = "PITest",
year = "Accessed: 2026",
url = {https://pitest.org}
}

@article{altmayer2025coverup,
  title={Coverup: Effective high coverage test generation for python},
  author={Altmayer Pizzorno, Juan and Berger, Emery D},
  journal={Proceedings of the ACM on Software Engineering},
  volume={2},
  number={FSE},
  pages={2897--2919},
  year={2025},
  publisher={ACM New York, NY, USA}
}

@article{gu2024testart,
  title={Testart: Improving llm-based unit testing via co-evolution of automated generation and repair iteration},
  author={Gu, Siqi and Zhang, Quanjun and Li, Kecheng and Fang, Chunrong and Tian, Fangyuan and Zhu, Liuchuan and Zhou, Jianyi and Chen, Zhenyu},
  journal={arXiv preprint arXiv:2408.03095},
  year={2024}
}

@article{dakhel2024effective,
  title={Effective test generation using pre-trained large language models and mutation testing},
  author={Dakhel, Arghavan Moradi and Nikanjam, Amin and Majdinasab, Vahid and Khomh, Foutse and Desmarais, Michel C},
  journal={Information and Software Technology},
  volume={171},
  pages={107468},
  year={2024},
  publisher={Elsevier}
}

@inproceedings{harman2025mutation,
  title={Mutation-guided llm-based test generation at meta},
  author={Harman, Mark and Ritchey, Jillian and Harper, Inna and Sengupta, Shubho and Mao, Ke and Gulati, Abhishek and Foster, Christopher and Robert, Herv{\'e}},
  booktitle={Proceedings of the 33rd ACM International Conference on the Foundations of Software Engineering},
  pages={180--191},
  year={2025}
}

@article{suppapitnarm2000simulated,
  title={A simulated annealing algorithm for multiobjective optimization},
  author={Suppapitnarm, A and Seffen, Keith A and Parks, Geoff T and Clarkson, PJ},
  journal={Engineering optimization},
  volume={33},
  number={1},
  pages={59--85},
  year={2000},
  publisher={Taylor \& Francis}
}

@article{panichella2017automated,
  title={Automated test case generation as a many-objective optimisation problem with dynamic selection of the targets},
  author={Panichella, Annibale and Kifetew, Fitsum Meshesha and Tonella, Paolo},
  journal={IEEE Transactions on Software Engineering},
  volume={44},
  number={2},
  pages={122--158},
  year={2017},
  publisher={IEEE}
}

@inproceedings{visser2004test,
  title={Test input generation with Java PathFinder},
  author={Visser, Willem and Pǎsǎreanu, Corina S and Khurshid, Sarfraz},
  booktitle={Proceedings of the 2004 ACM SIGSOFT international symposium on Software testing and analysis},
  pages={97--107},
  year={2004}
}

@article{tufano2020unit,
  title={Unit test case generation with transformers and focal context},
  author={Tufano, Michele and Drain, Dawn and Svyatkovskiy, Alexey and Deng, Shao Kun and Sundaresan, Neel},
  journal={arXiv preprint arXiv:2009.05617},
  year={2020}
}

@inproceedings{chen2024chatunitest,
  title={Chatunitest: A framework for llm-based test generation},
  author={Chen, Yinghao and Hu, Zehao and Zhi, Chen and Han, Junxiao and Deng, Shuiguang and Yin, Jianwei},
  booktitle={Companion Proceedings of the 32nd ACM International Conference on the Foundations of Software Engineering},
  pages={572--576},
  year={2024}
}

@inproceedings{sapozhnikov2024testspark,
  title={Testspark: Intellij idea's ultimate test generation companion},
  author={Sapozhnikov, Arkadii and Olsthoorn, Mitchell and Panichella, Annibale and Kovalenko, Vladimir and Derakhshanfar, Pouria},
  booktitle={Proceedings of the 2024 IEEE/ACM 46th international conference on software engineering: companion proceedings},
  pages={30--34},
  year={2024}
}

@article{shin2024retrieval,
  title={Retrieval-augmented test generation: How far are we?},
  author={Shin, Jiho and Harzevili, Nima Shiri and Aleithan, Reem and Hemmati, Hadi and Wang, Song},
  journal={arXiv preprint arXiv:2409.12682},
  year={2024}
}

@inproceedings{alshahwan2024automated,
  title={Automated unit test improvement using large language models at meta},
  author={Alshahwan, Nadia and Chheda, Jubin and Finogenova, Anastasia and Gokkaya, Beliz and Harman, Mark and Harper, Inna and Marginean, Alexandru and Sengupta, Shubho and Wang, Eddy},
  booktitle={Companion Proceedings of the 32nd ACM International Conference on the Foundations of Software Engineering},
  pages={185--196},
  year={2024}
}

@inproceedings{lemieux2023codamosa,
  title={Codamosa: Escaping coverage plateaus in test generation with pre-trained large language models},
  author={Lemieux, Caroline and Inala, Jeevana Priya and Lahiri, Shuvendu K and Sen, Siddhartha},
  booktitle={2023 IEEE/ACM 45th International Conference on Software Engineering (ICSE)},
  pages={919--931},
  year={2023},
  organization={IEEE}
}

@misc{lizard,
  title = {Lizard: A Code Complexity Analyzer},
  howpublished = {\url{https://github.com/terryyin/lizard}},
  note = {Accessed 2026-06-23}
}

@misc{vllm,
  title = {vllm: A High-Performance LLM Inference and Serving Framework},
  howpublished = {\url{https://vllm.ai}},
  note = {Accessed 2024-06-23}
}

@inproceedings{lin2021graph,
  title={Graph-based seed object synthesis for search-based unit testing},
  author={Lin, Yun and Ong, You Sheng and Sun, Jun and Fraser, Gordon and Dong, Jin Song},
  booktitle={Proceedings of the 29th ACM Joint Meeting on European Software Engineering Conference and Symposium on the Foundations of Software Engineering},
  pages={1068--1080},
  year={2021}
}

@misc{jquickcurl,
  author       = {{paohaijiao}},
  title        = {jquick-curl: A Java wrapper for libcurl},
  howpublished = {\url{https://github.com/paohaijiao/jquick-curl}},
  note = {Accessed 2024-06-23}
}

@misc{adkjava,
  author       = {{Google}},
  title        = {Agent Development Kit (ADK) for Java},
  howpublished = {\url{https://github.com/google/adk-java}},
  note = {Accessed 2024-06-23}
}

@misc{a2a,
  author       = {{A2A Project}},
  title        = {A2A: An Open-Source Framework for Building Autonomous Agents in Java},
  howpublished = {\url{https://github.com/a2aproject/a2a-java}},
  note         = {Accessed 2024-06-23}
}

@misc{ascope,
  author       = {{agentscope-ai}},
  title        = {Build Production-Ready AI Agents in Java},
  howpublished = {\url{https://github.com/agentscope-ai/agentscope-java}},
  note         = {Accessed 2024-06-23}
}

@article{ccpaper,
  title={Evaluating the Dependency Between Cyclomatic Complexity and Response For Class},
  author={Stavtsev, Maxim and Bugayenko, Yegor},
  journal={arXiv preprint arXiv:2410.06416},
  year={2024}
}

@article{xu2026llm,
  title={LLM-based Mockless Unit Test Generation for Java},
  author={Xu, Qinghua and Wang, Guancheng and Briand, Lionel and Guo, Zhaoqiang and Liu, Kui},
  journal={arXiv preprint arXiv:2605.26851},
  year={2026}
}

@misc{fireworkAI,
  title = {Fireworks AI},
  howpublished = {\url{https://fireworks.ai/models/fireworks/deepseek-v4-flash}},
  note = {Accessed 2024-06-23}
}

@misc{codex_cli,
  author       = {{OpenAI}},
  title        = {{Codex CLI}},
  howpublished = {\url{https://github.com/openai/codex}},
  year         = {2026},
  note         = {Accessed: 2026-10-27}
}

@misc{codex_custom_provider,
  author       = {{gptzzz}},
  title        = {{Codex Custom Provider}},
  howpublished = {\url{https://github.com/gptzzz/codex-custom-provider}},
  year         = {2026},
  note         = {Accessed: 2026-10-27}
}

@misc{claude_code,
  author       = {{Anthropic}},
  title        = {{Claude Code}},
  howpublished = {\url{https://github.com/anthropics/claude-code}},
  year         = {2026},
  note         = {Accessed: 2026-10-27}
}

@misc{claude_code_proxy,
  author       = {{suparious}},
  title        = {{Claude Code Proxy}},
  howpublished = {\url{https://github.com/suparious/claude-code-proxy}},
  year         = {2026},
  note         = {Accessed: 2026-10-27}
}

@techreport{nist2024genai,
  author      = {Autio, Chloe and Schwartz, Reva and
                 Dunietz, Jesse and Jain, Shomik and
                 Stanley, Martin and Tabassi, Elham and
                 Hall, Patrick and Roberts, Kamie},
  title       = {Artificial Intelligence Risk Management Framework:
                 Generative Artificial Intelligence Profile},
  institution = {National Institute of Standards and Technology},
  number      = {NIST AI 600-1},
  year        = {2024},
  doi         = {10.6028/NIST.AI.600-1}
}

@misc{owasp2025llm,
  author       = {{OWASP Foundation}},
  title        = {{OWASP Top 10 for Large Language Model Applications}},
  year         = {2025},
  howpublished = {\url{https://genai.owasp.org/llm-top-10/}},
  note         = {Accessed: 2026-10-27}
}

@inproceedings{sallou2024breaking,
  author    = {Sallou, June and Durieux, Thomas and
               Panichella, Annibale},
  title     = {Breaking the Silence: The Threats of Using
               {LLMs} in Software Engineering},
  booktitle = {Proceedings of the 46th IEEE/ACM International
               Conference on Software Engineering:
               New Ideas and Emerging Results (ICSE-NIER)},
  year      = {2024},
  doi       = {10.1145/3639476.3639764}
}

@misc{github2026copilotbudget,
  author       = {{GitHub}},
  title        = {Budgets for Usage-Based Billing},
  year         = {2026},
  howpublished = {\url{https://docs.github.com/en/copilot/concepts/billing-and-usage/organizations-and-enterprises/budgets}},
  note         = {GitHub Copilot documentation.
                  Accessed: 2026-10-28}
}

@misc{anthropic2026enterpriseusage,
  author       = {{Anthropic}},
  title        = {Claude Enterprise Consumption Guide},
  year         = {2026},
  howpublished = {\url{https://support.claude.com/en/articles/14782391-claude-enterprise-consumption-guide}},
  note         = {Accessed: 2026-10-28}
}

@misc{openai2026codexconfig,
  author       = {{OpenAI}},
  title        = {Codex: Advanced Configuration},
  year         = {2026},
  howpublished = {\url{https://learn.chatgpt.com/docs/config-file/config-advanced}},
  note         = {Accessed: 2026-11-01}
}

@misc{anthropic2026gateway,
  author       = {{Anthropic}},
  title        = {Claude Code: LLM Gateway Configuration},
  year         = {2026},
  howpublished = {\url{https://docs.anthropic.com/en/docs/claude-code/llm-gateway}},
  note         = {Accessed: 2026-11-01}
}

@misc{klemmer2024aiassistants,
  author       = {Jan H. Klemmer and others},
  title        = {Using AI Assistants in Software Development:
                  A Qualitative Study on Security Practices
                  and Concerns},
  year         = {2024},
  eprint       = {2405.06371},
  archivePrefix = {arXiv},
  primaryClass = {cs.SE},
  url          = {https://arxiv.org/abs/2405.06371}
}

@misc{snyk2025aicode,
  author       = {{Snyk}},
  title        = {AI Code Security Report},
  year         = {2025},
  howpublished = {\url{https://go.snyk.io/rs/677-THP-415/images/AI%20Report%20-ENG.pdf?version=0}},
  note         = {Accessed: 2026-11-01}
}

\end{document}